\documentclass[8pt,aps,showpacs,nofootinbib,preprint]{revtex4-1}

\usepackage{hyperref}
\usepackage{latexsym}
\usepackage{epsf}
\usepackage{epsfig}
\usepackage{amssymb,amsmath,amsfonts,amsxtra,amsthm}
\usepackage{bm}
\usepackage{dcolumn}
\usepackage{array}
\usepackage{tabularx}
\usepackage{enumerate}
\usepackage{hhline}
\usepackage{subfigure}
\usepackage{color}
\usepackage{graphics}
\newcommand{\beq}{\begin{equation}}
\newcommand{\eeq}{\end{equation}}
\newcommand{\beqa}{\begin{eqnarray}}
\newcommand{\eeqa}{\end{eqnarray}}
\newcommand{\barr}{\begin{array}}
\newcommand{\earr}{\end{array}}

\begin{document}


\title{Hybrid Quasi-Single Field Inflation}
\date{\today}

\author{Yi Wang$^{a,b}$}
\author{Yi-Peng Wu$^c$} \email[Electronic address: ]{ypwu@resceu.s.u-tokyo.ac.jp}
\author{Jun'ichi Yokoyama$^{c,d,e}$}
\author{Siyi Zhou$^{a,b}$}
\affiliation{${}^a$Department of Physics, The Hong Kong University of Science and Technology, \\
	Clear Water Bay, Kowloon, Hong Kong, P.R.China}
\affiliation{${}^b$Jockey Club Institute for Advanced Study, The Hong Kong University of Science and Technology, \\
	Clear Water Bay, Kowloon, Hong Kong, P.R.China}
\affiliation{${}^c$Research Center for the Early Universe (RESCEU),
	Graduate School of Science, The University of Tokyo,
	Tokyo 113-0033, Japan}

\affiliation{${}^d$Department of Physics,
	Graduate School of Science, \\ The University of Tokyo,
	Tokyo 113-0033, Japan}
\affiliation{${}^e$Kavli Institute for the Physics and Mathematics of the 
	universe (Kavli IPMU), WPI, UTIAS,\\The University of Tokyo, 5-1-5 Kashiwanoha, Kashiwa 277-8583, Japan}

\begin{abstract}
The decay of massive particles during inflation generates characteristic signals in the squeezed limit of the primordial bispectrum. 
These signals are in particular distinctive in the regime of the quasi-single field inflation, where particles are oscillating with masses comparable to the Hubble scale.
We apply the investigation to a class of scalar particles that experience a so-called waterfall phase transition in the isocurvature direction 
driven by the symmetry-breaking mechanism based on the hybrid inflation scenario. With a time-varying mass,
a novel shape of oscillatory bispectrum is presented as the signature of a waterfall phase transition during inflation.
\end{abstract}

\preprint{RESCEU-05/18}
\maketitle

\section{Introduction}

Inflation \cite{Starobinsky:1980te,Sato:2015dga} has been a leading paradigm for the primordial universe 
and most likely an observable phenomenon triggered at the highest energy scale among all current experiments.
Observations for the inflationary parameters are in well agreement with the simplest scenario where 
only a single scalar degree of freedom is involved. The primordial scalar mode, $\zeta$, seems to be responsible to the
soft breaking of the global time-translation symmetry during inflation \cite{Cheung:2007st,Baumann:2014nda}. 
One therefore expects that the correlation functions of $\zeta$
follow the single-field consistency relations \cite{Maldacena:2002vr,Creminelli:2004yq}, 
and signals from the primordial bispectrum vanishes in the squeezed limit unless there exist new particles that decay into $\zeta$. 
This is a window opened known as the cosmological collider physics \cite{Arkani-Hamed:2015bza}, 
see also \cite{Chen:2009we,Chen:2009zp,Sefusatti:2012ye,Norena:2012yi,Assassi:2012zq,Dimastrogiovanni:2015pla,Schmidt:2015xka,Meerburg:2016zdz,MoradinezhadDizgah:2018ssw}.   

Massive particles with a mass $m$ around the order of the Hubble parameter $H$ during inflation appear to be the main target
for the cosmological collider research. This is because massive fields with $m \gg H$ are usually recast into an effective background,
while particles with $m \ll H$ often lead to generalized version of inflationary dynamics beyond the single-field configuration.
One of the characteristic signature of massive fields with $m \sim H$ comes from the three-point function of $\zeta$ in the squeezed limit.
In particular, an oscillatory behavior of the non-Gaussian correlation functions can be generated by the decay of massive fields into $\zeta$ with $m > 3H/2$,
as a consequence of the quantum interference between massless and massive mode functions.
This type of oscillatory feature in the squeezed non-Gaussianities can be easily derived from the spatial dilation symmetry of inflation \cite{Assassi:2012zq}. 

Quasi-single field inflation\footnote{The effective single-field inflation reproduced from a multi-field configuration with heavy field masses $m \gg H$ was first investigated in \cite{Yamaguchi:2005qm}.}
 \cite{Chen:2009we,Chen:2009zp} is a pioneer study which resolves the imprints of massive fields in the non-Gaussianities with $m \geq H$.
As a practical example, one can consider an inflaton field slowly rolling with a turning trajectory, whose kinetic term is naturally mixed with a massive mode
in the isocurvature direction. In this scenario, the isocurvature perturbations receive an effective mass from both the potential and 
the kinetic couplings to the inflaton. Recent studies \cite{An:2017hlx,An:2017rwo,Chen:2015lza,Chen:2017ryl} have confirmed that
the bispectrum involved with the decay of the massive field are indeed oscillating with respect to the ratio $k_{\rm long}/k_{\rm short}$,
where $k_{\rm long}$ ($k_{\rm short}$) is a larger (smaller) scale mode which exits the horizon earlier (later) in time.
The amplitude of such an oscillation is enhanced in the squeezed limit ($k_{\rm long}/k_{\rm short} \rightarrow 0$) if the
large effective mass comes from the field potential \cite{Chen:2015dga,Chen:2017ryl}, 
and it is enhanced in the equilateral limit ($k_{\rm long}/k_{\rm short} \rightarrow 1$)  
if the large effective mass is induced by the kinetic coupling \cite{An:2017hlx,An:2017rwo}.
These oscillatory signals are usually small due to the Boltzmann suppression factor $e^{- \pi m/H}$. There are a few possibilities to generate this type of oscillatory signals without any exponential suppression. For example, from classical oscillations of inflationary fluctuations from a sharp feature (the classical primordial standard clock) \cite{Chen:2011zf,Saito:2012pd,Chen:2014cwa},  particle production enhanced by inflaton kinetic energy during monodromy inflation \cite{Flauger:2016idt}, and 
production of heavy particles due to finite-temperature effects during warm inflation \cite{Tong:2018tqf}. 

It is also interesting to seek for signals from another inflationary scenario involving massive fields. 
For example, in the hybrid inflation scenario \cite{Linde:1993cn} 
an isocurvature field acquires an effective mass by virtue of a direct coupling to the inflaton field. 
This effective mass square decays with the slow rolling of the inflaton and becomes negative after some critical points where
 the isocurvature field can start to roll down into a new minimum of the potential (namely a waterfall phase transition)
 \footnote{Phase transitions of an isocurvature field can also be realized by its direct coupling with gravity, see \cite{Nagasawa:1991zr,Yokoyama:1989pa}.}.
If the negative mass term during the falling down process is sufficiently large and 
the massive field comes to dominate the energy density, the slow-roll inflation will be terminated soon after the waterfall phase transition
 \cite{Linde:1993cn}. 
On the other hand, if this mass is not much greater than the Hubble parameter, this field will induce a secondary slow-roll inflation
along the direction orthogonal to the original inflation \cite{Clesse:2010iz,GarciaBellido:1996qt}, in a way similar to the double-inflation scenario \cite{Yamaguchi:2003fp,Yamaguchi:2004tn,Kawasaki:2006zv}.
These features lead us the curiosity on the implications of such a kind of waterfall field to the primordial spectra.
We are especially interested in the case where the waterfall field is always subdominant in the total energy density, even after the phase transition.
In this case the background dynamics is still governed by the single-field inflation paradigm, 
yet once the waterfall field falls down into the new minimum and acquires a large
positive mass, the decay of isocurvature field perturbations can generate oscillating features in the bispectum of $\zeta$ through the
same mechanism as that of the quasi-single field inflation \cite{Chen:2009zp,Arkani-Hamed:2015bza}.

\begin{figure}
	\begin{center}
		\includegraphics[width=70mm]{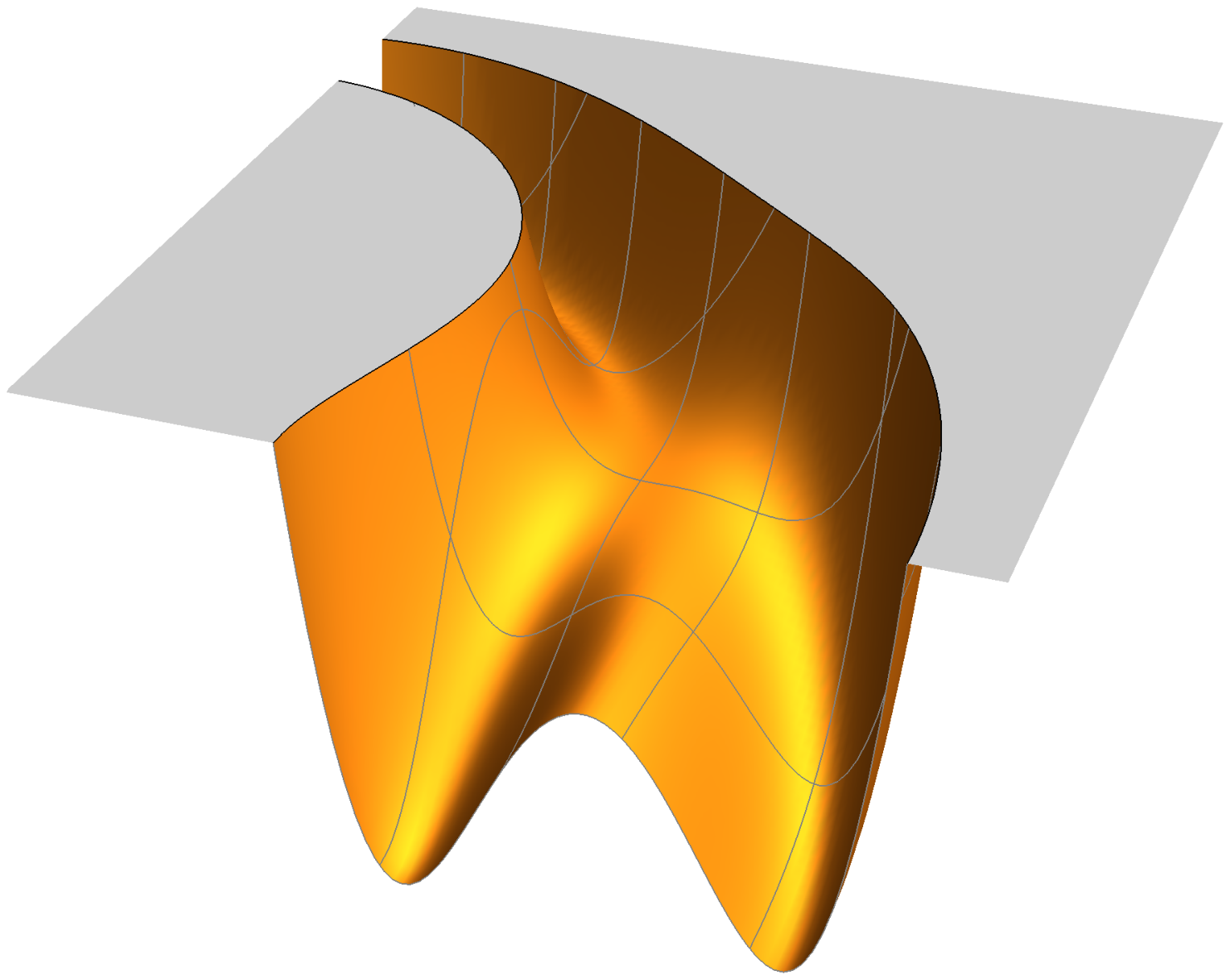}\qquad\qquad\qquad
		\includegraphics[width=60mm]{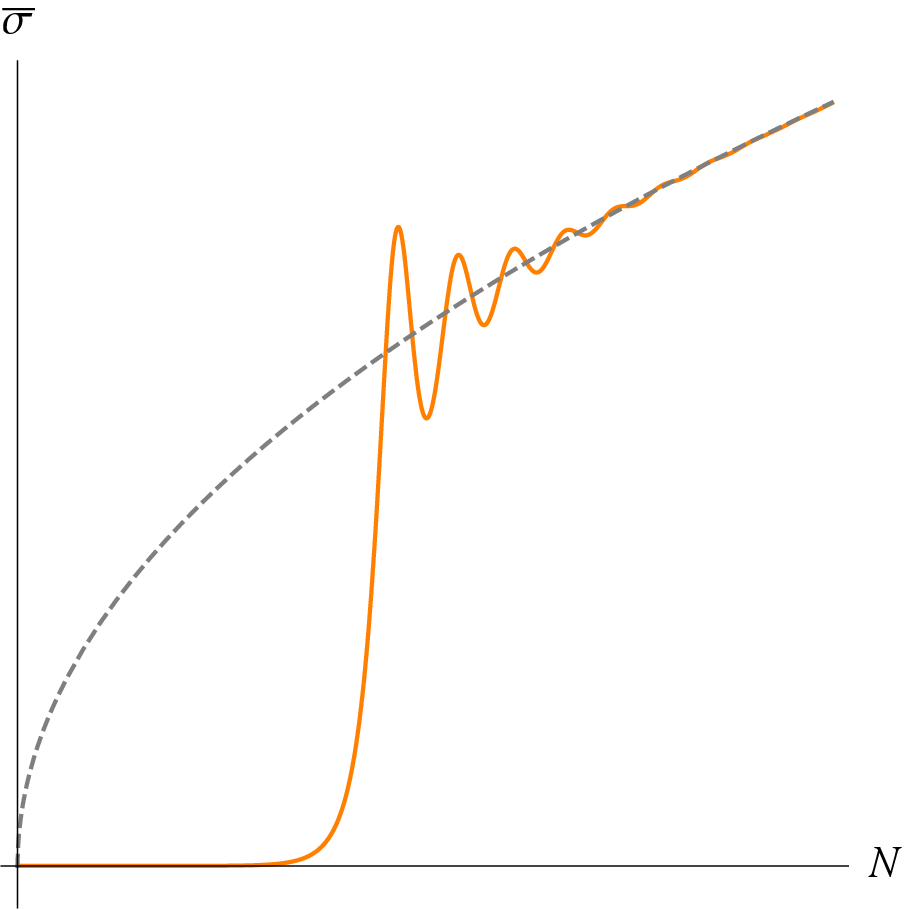}
	\end{center}
	\caption{\label{fig 1}Left panel: Illustration of the hybrid-inflation potential with a turning minimum. 
		Right panel: A numerical example for $\bar{\sigma}(N)$ that generates an ocsillating trajectory due to the waterfall phase transition.
		The dashed line depicts the evolution of the positive branch of $\bar{\sigma}_{\rm min}$.
	}
\end{figure}

In this work, we consider the presence of a waterfall phase transition in a turning trajectory 
by introducing a coupling as $\phi^2 \sigma^2/2$
between the inflaton $\phi$ and the waterfalling isocurvature field $\sigma$
 from the hybrid inflation \cite{Linde:1993cn} to the simplest
quasi-single field inflation model with a constant turning radius \cite{Chen:2009zp}. 
The hybrid potential of our model can be schematically illustrated by Fig. \ref{fig 1}.
We show that both the hybrid inflation and the constant turn inflation scenario get benefits from each other as follows:

\begin{itemize}
	\item A turning trajectory in the inflationary direction naturally breaks the degeneracy of the local-minimal states in the isocurvature direction,
	even if these minimal states share the same vacuum expectation value. This is true when the waterfall field is accelerated by a non-vanishing
	centrifugal force due to the inflaton velocity. For example, in the hybrid inflation \cite{Linde:1993cn} with a double-well potential $\propto (\sigma^2 - v^2)^2$ 
	the rate of $\sigma$ to fall down into the minima at $\pm v$ now depends on the initial conditions at the time of phase transition.
	As a result, the rate of multi-stream inflation with domain wall formation \cite{Li:2009sp,Afshordi:2010wn,Liu:2015dda} in this scenario can be controlled by the model parameters.
	
	\item The oscillating signature in the $\zeta$-bispectrum can be efficiently amplified in the squeezed limit
	$(k_{\rm long}/k_{\rm short} \rightarrow 0)$, when comparing with the results of the constant-turn inflation \cite{An:2017hlx,An:2017rwo}.
	This is due to the fact that a waterfall phase transition driven by the unstable part of the potential along the isocurvature direction in general leads to 
	a period with tachyonic instability where isocurvature field perturbations are growing on superhorizon scales. 
	
	
	\item A waterfall phase transition can act as a trigger for classical oscillations of the isocurvature field (see Fig. \ref{fig 1}).
	The small oscillations of the isocurvature field can result in observable features that are sensitive to the background expansion rate of the primordial universe.
	These oscillating features in the primordial spectra are recognized as the classical standard clock signals, 
	which are useful for testing inflation and its alternative scenarios \cite{Chen:2011zf,Saito:2012pd,Chen:2014cwa}.
\end{itemize}

This paper is organized as the follows. In Section~\ref{modelhybrid}, we introduce a model which incorporates 
the constant turn inflation with the waterfall potential from the hybrid inflation scenario. We study the classical field evolution via the stochastic approach in Section~\ref{classical field evolutions}.
In Section~\ref{quantumfluctuation}, we outline our approach for the quantum field perturbations in this scenario, and
the numerical results for the two-point and three-point correlation functions are given in Section~\ref{Sec:Imprints}. 
Finally, we give summary and discussions in Section~\ref{Sec:discussion}.

\section{Hybrid inflation with a turn}\label{modelhybrid}

We construct a theory based on the quasi-single field inflation \cite{Chen:2009zp} with a turning trajectory.
It is convenient to apply the polar-coordinate representation to decompose the inflaton $\phi$
by the tangential mode $\theta = \phi/R$ and the isocurvature field (or the isocurvaton) 
by the radial mode $\sigma$. The action is given by
\begin{equation}\label{model}
	S_{qs}=\int d^4 x \sqrt{-g} \left[ -\frac{1}{2}(R+\sigma-\sigma_0^\ast)^2  (\partial \theta)^2 -\frac{1}{2}(\partial\sigma)^2 - V(\theta, \sigma) \right] ,
\end{equation}
where $R$ and $\sigma_0^\ast$ are constants with the same dimension of the mass. 
We can rephrase the action by the usual representation of the inflaton field $\phi = R \theta$,
which has the same dimension as $\sigma$, and \eqref{model} is rewritten as
\begin{equation}
S_{qs}=\int d^4 x \sqrt{-g} \left[ -\frac{1}{2}(1+\frac{\sigma-\sigma_0^\ast}{R})^2  (\partial \phi)^2 -\frac{1}{2}(\partial\sigma)^2 - V(\phi, \sigma) \right].
\end{equation}
The kinetic coupling $(\sigma-\sigma_0^\ast)^2(\partial \phi)^2/R^2$ is assumed to be perturbatively small.

Both $\theta$ and $\sigma$ exhibit a well-defined coherent part (or the zero-mode part) $\theta_0$ and $\sigma_0$,
and their evolution in the spatially homogeneous background are governed by \cite{Chen:2015dga}:
\begin{align} \label{eq:Friedmann_1}
3M_p^2H^2 &= \frac{1}{2} (\tilde{R} + \sigma_0)^2 \dot{\theta}^2_0 +\frac{1}{2}\dot{\sigma}_0^2 +V(\theta_0, \sigma_0), \\ \label{eq:Friedmann_2}
-2M_p^2 \dot{H} &= (\tilde{R} + \sigma_0)^2 \dot{\theta}^2_0 + \dot{\sigma}_0^2, \\
(\tilde{R} + \sigma_0)^2\ddot{\theta}_0 +3 (\tilde{R} + \sigma_0)^2 H\dot{\theta}_0 + V_\theta &= 0, \\
\ddot{\sigma}_0 + 3H \dot{\sigma}_0 + V_\sigma &= (\tilde{R} + \sigma_0) \dot{\theta}_0^2, \label{eom:sigma0}
\end{align}
where $\tilde{R}\equiv R - \sigma_0^\ast$, $V_\theta \equiv \partial V/\partial \theta$ ($V_\sigma \equiv \partial V/\partial \sigma$), 
and $(\tilde{R} + \sigma_0) \dot{\theta}_0^2$ features the centrifugal force applied to the isocurvaton $\sigma$ 
(which is nothing but the waterfall field in the current study).
The time evolution of $\sigma_0$ plays a fundamental role in our scenario, but we shall keep 
$\sigma_0 \ll \tilde{R}$ and $\dot{\sigma}_0^2 \ll R^2\dot{\theta}^2_0$ so that $\theta_0$ dominates the inflationary dynamics
and observables.

A phase transition of $\sigma_0$ from the initial expectation value $\sigma_0^\ast$ 
may be realized by direct couplings between $\theta$ and $\sigma$ in the potential.
As a simple example, we consider the symmetry-breaking scenario driven by the rolling of $\theta$ as \cite{Linde:1993cn}
\begin{equation}
V(\theta, \sigma) = V_{\rm sr}(\theta) + \frac{\lambda}{4}\left[ (\sigma-\sigma_0^\ast)^2 - v^2\right]^2 + \frac{g^2}{2}\theta^2(\sigma-\sigma_0^\ast)^2,
\end{equation}
where $V_{\rm sr}(\theta)$ stands for the potential of the slow-roll inflation and $\lambda$ is a dimensionless parameter. 
$g$ and $v$ are parameters with the same dimension of the mass. In terms of $\phi = R\theta$, one can identify the dimensionless coupling constant $\bar{g} = g/R$
from the interaction term $g^2\theta^2(\sigma-\sigma_0^\ast)^2 \equiv \bar{g}^2\phi^2(\sigma-\sigma_0^\ast)^2$.
The critical value of phase transition in the isocurvature direction is $\theta_c^2 = \lambda v^2/g^2$ when $\sigma = \sigma_0^\ast$.
There are two interesting regimes in the theory:
\begin{enumerate}
	\item If $V_{\rm sr}(\theta_c) \ll \lambda v^4/4$, the waterfall field comes to dominate the energy density of the universe after phase transition,
	and the slow-roll inflation due to $V_{\rm sr}(\theta)$ can be terminated by a rapid waterfall of $\sigma$  \cite{Linde:1993cn} or be replaced by a secondary inflation
	driven by the slow rolling of $\sigma$ \cite{GarciaBellido:1996qt}, depending on the value of the critical mass
	$m_{ c}^2 \equiv g^2\theta_c^2 = \lambda v^2$. We consider this regime as the usual hybrid inflation scenario.
	
	\item If $V_{\rm sr}(\theta_c) \gg \lambda v^4/4$, the slow-roll inflation due to $\theta$ continues after the phase transition,
	but the inflationary trajectory is shifted by the rolling of the isocurvaton in the radial direction. 
	This regime can be cast into a broader class of the quasi-single field inflation \cite{Chen:2009zp},
	provided that $m_{ c}^2 \sim H^2$. 
	If $m_{ c}^2 \gg H^2$, namely $\sigma$ acquires a negative effective mass much greater than the hubble parameter, 
	the transition of $\sigma$ finishes instantly and the predictions of the theory converge to that of the single-field inflation.
\end{enumerate}
We therefore restrict the current study to the second regime.

\subsection{Inflation with a constant turn}

Let us put $\sigma$ very close to $\sigma_0^\ast$ at the beginning.
When $\theta \gg \theta_c$, its mass reads $m_\sigma^2 \approx g^2\theta^2 \gg m_c^2$. 
If we consider that $m_c^2 \gtrsim \mathcal{O}(H^2)$, then $\sigma_0$ remains very close to $\sigma_0^\ast$
before the phase transition. Thus we can expand $V_\sigma$ around the minimum as
\begin{align}\label{V_sigma expansion}
V_\sigma (\sigma) = V_{\sigma\sigma}(\sigma_0^\ast) (\sigma -\sigma_0^\ast) 
+ \frac{1}{2} V_{\sigma\sigma\sigma}(\sigma_0^\ast) (\sigma -\sigma_0^\ast)^2+\cdots, 
\end{align}
where $V_{\sigma\sigma}(\sigma_0^\ast) = g^2\theta^2 -\lambda v^2 \gg m_{ c}^2$ at this stage.
The equation of motion for $\sigma_0$ at zeroth-order reads
\begin{align}\label{eom:sigma0_zeroth}
\ddot{\sigma}_0 + 3H\dot{\sigma}_0 = R \dot{\theta}_0^2,
\end{align}
since $V_\sigma(\sigma_0^\ast) =0$.

To see the effect of a non-vanishing centrifugal force, it is convenient to consider a constant inflaton velocity $\dot{\theta}_0$.
We will show that this condition holds in a generic class of slow-roll inflation models.
Ignoring for the moment the mild time-evolution of the Hubble parameter, the solution of \eqref{eom:sigma0_zeroth} takes the form of
\begin{align}
\sigma_0 = \sigma_0^\ast + \frac{R \dot{\theta}_0^2}{3H} t - \frac{c_0}{3H}e^{-3Ht},
\end{align}
where $c_0$ can be determined by a given initial condition but its contribution is suppressed by the exponential factor.
One can check that higher-order corrections to $\sigma_0$ due to \eqref{V_sigma expansion} are also decaying with time.
As a result, a non-zero centrifugal force leads to a deviation from the minimum as 
$\sigma_0 - \sigma_0^\ast \propto \Delta t$ after some finite duration. 

\subsection{Dynamics of waterfall phase transition}

To make a concrete discussion on the dynamics of phase transition, we redefine the isocurvaton field into a dimensionless parameter 
as $\bar{\sigma} \equiv (\sigma_0 - \sigma_0^\ast)/R$
and we rewrite the equation of motion \eqref{eom:sigma0} with respect to the number of e-folds $N =\ln a$ as 
\begin{align}\label{eom:sigma_bar}
\frac{d^2 \bar{\sigma}}{d N^2} + (3-\epsilon_H)\frac{d\bar{\sigma}}{d N} + 
\left[g^2\frac{\theta_0^2}{H^2} + \frac{\lambda}{H^2}(R^2\bar{\sigma}^2 -v^2)\right] \bar{\sigma} =  \frac{\dot{\theta}_0^2}{H^2},
\end{align}
where $\epsilon_H = -\dot{H}/H^2$ is the first slow-roll parameter.
We also specify the slow-roll potential by the chaotic inflation 
form
\begin{align}\label{V_choatic}
V_{\rm sr}(\theta) = \mu^4 \theta^p,
\end{align}
which recovers the prototype scenario \cite{Linde:1993cn} when taking $p =2$.
In the single-field limit both slow-roll parameters $\epsilon_H$ and $\eta_H = \dot{\epsilon_H}/(H\epsilon_H)$ beocme independent of $\mu$. 
To sustain a large enough e-folding number for a slow-roll inflation one requires $R \gg M_p$,
and the inflaton solution is well-approximated by 
\begin{align}\label{sol:theta_0}
\theta_0 = \theta_c -p\frac{M_p^2}{R^2 \theta_c^2}(N -N_c),
\end{align}
throughout our discussion, where $N_c$ is the epoch when $\theta = \theta_c$.
The potential \eqref{V_choatic} supports a constant centrifugal force $\dot{\theta}/H = -p M_p^2/(R^2 \theta_c^2)$.

Without a loss of generality, we choose $N_c =0 $ and approximate the value of $\bar{\sigma}$ at the critical point as
$\bar{\sigma}(N_c) = \bar{\sigma}_c \approx \frac{\dot{\theta}_0^2}{3H_c^2}\Delta N$, 
where $H_c^2 \equiv 3M_p^2/(\mu^4 \theta_c^p)$ and $\Delta N$ is the e-folding number before phase transition. 
Assuming that $\bar{\sigma}$ starts to roll down smoothly once $\theta$ reaches the critical value, we can
simplify the equation of motion \eqref{eom:sigma_bar} as
\begin{align}\label{eom:sigma_bar simplify}
\frac{d\bar{\sigma}}{d N} -A N \bar{\sigma} = B,
\end{align} 
where we introduce two useful parameters as
\begin{align}
A = \frac{2 p g^2}{3 H_c^2\theta_c} \frac{M_p^2}{R^2}, \qquad B = \frac{p^2 M_p^4}{3 R^4 \theta_c^4}.
\end{align}
The solution of \eqref{eom:sigma_bar simplify} reads
\begin{align}\label{sol:sigma_bar}
\bar{\sigma} = e^{AN^2/2 }\left[ \bar{\sigma}_c +B \sqrt{\frac{\pi}{2A}} \, \mathrm{Erf}\left(\sqrt{\frac{A}{2}}N\right) \right].
\end{align}
Since $\bar{\sigma}_c \approx B \Delta N$, it turns out that $\bar{\sigma} \simeq \bar{\sigma}_c e^{AN^2/2 }$ if $\Delta N \gg \sqrt{\pi/(2A)}$.

\begin{figure} 
	\begin{center}
		\includegraphics[width=70mm]{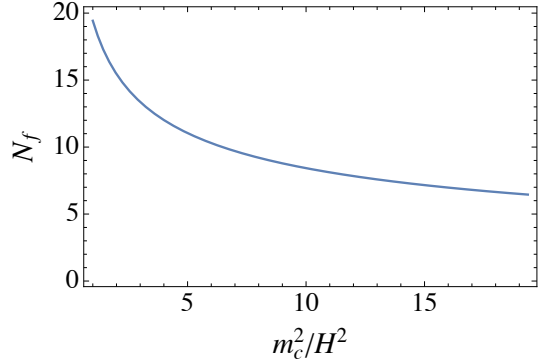}
		
	\end{center}
	\caption{\label{Fig2}
		The duration of the waterfall field to reach the positive branch of $\bar{\sigma}_{\rm min}$ since $N = 0$ with respect to the critical mass $m_c^2 = g^2\theta_c^2 = \lambda v^2$.
	}
\end{figure}

The waterfall ends when $\sigma$ reaches the new minima at $\theta > \theta_c$. 
Taking the solution \eqref{sol:theta_0}, the condition $V_\sigma = 0$ with $N > 0$ indicates that
\begin{align}\label{sigma_minima}
\bar{\sigma}_{\rm min} = \pm \sqrt{\frac{3H_c^2 A}{\lambda R^2} N},
\end{align}
which is evolving with time. We can estimate the epoch $N = N_f$ at the end of the phase transition 
by $\bar{\sigma}(N_f) = \bar{\sigma}_{\rm min}$. The result from the solution \eqref{sol:sigma_bar} is given in Fig. \ref{Fig2}.

\section{Classical field evolution}\label{classical field evolutions}

\begin{figure}
	\begin{center}
		\includegraphics[width=110mm]{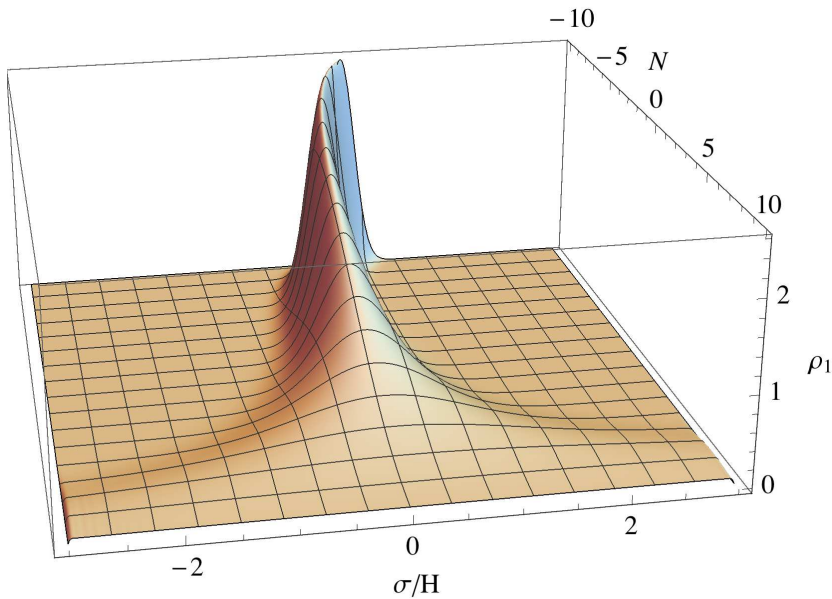}
	\end{center}
	\caption{\label{fig_standard_PDF}The evolution of the one-point probability distribution function $\rho_1$ of the waterfall field without centrifugal force ($\xi = 0$). The waterfall phase transition starts at $N =0$.
		The initial value $\sigma_0 = 0.5 H$ is used.
		The potential parameters are fixed by $g/H = 4$, $\lambda = 10^{-6}$, $\theta_c = 1$, $p =1$ , $R/M_p = 10$ and $H/ M_p = 10^{-5}$.
	}
\end{figure}

In this section we investigate the behavior of the long wavelength part of the waterfall field $\sigma$.
By imposing a classical initial value $\sigma_0$ and a certain probability distribution of the long wavelength modes, we study the evolution of such a distribution undergoing the waterfall phase transition.
In general, we identify the peak of the probability distribution with the classical expectation value $\sigma_0(t)$ at a given time $t$.
The decomposition $\sigma \rightarrow \sigma_0 + \delta\sigma$, with $\delta\sigma$ as quantum fluctuations, becomes well-defined if the spread-out (or decoherent) time-scale of the distribution is comparable to the duration of inflation.

The long wavelength part of $\sigma$ can be treated as an auxiliary classical field, whose one-point probability distribution function (PDF), $\rho_1[\sigma(\mathbf{x}, t)]$, satisfies the Fokker-Planck equation
\cite{Starobinsky:1986,Starobinsky:1994bd}
\begin{equation}\label{FP eq1}
\frac{\partial\rho_1[\sigma]}{\partial t} =
\frac{1}{3H} \frac{\partial}{\partial\sigma} 
\left( \rho_1[\sigma]\, \frac{\partial V_{\rm eff}}{\partial\sigma}  \right) +
\frac{H^3}{8\pi^2}
\frac{\partial^2\rho_1[\sigma]}{\partial\sigma^2},
\end{equation}
where the effective field potential $V_{\rm eff}$ deduced from the classical equation of motion \eqref{eom:sigma0} reads
\begin{align}
V_{\rm eff} = \frac{1}{2}(g^2\theta_0^2 - \lambda v^2 - \xi \dot{\theta}_0^2) \sigma^2 + \frac{\lambda}{4} \sigma^4 - \xi R\dot{\theta}_0^2 \sigma.
\end{align}
Here the terms with $\dot{\theta}_0^2$ arise due to the centrifugal force of the turning trajectory, and we have moved the origin to $\sigma = \sigma_0^\ast = 0$ for convenience.   
We introduce a new parameter $\xi$ to control the magnitude of the centrifugal force only for the study in this section, and the model \eqref{model} can be recovered by taking $\xi = 1$.
 
We adopt the analytical expression \eqref{sol:theta_0} for $\theta_0$ and solve $\rho_1$ numerically by \eqref{FP eq1}.
We impose an initial distribution $\rho_1(t_0)\equiv \rho_{10}$ of the form
\begin{align}
\rho_{10}= \frac{H}{\sqrt{2\pi \langle \sigma^2\rangle_0}} \exp\left[\frac{-(\sigma-\sigma_0)^2}{2\sqrt{\langle \sigma^2\rangle_0}}\right],
\end{align}
where the initial variance $\langle \sigma^2\rangle_0$ is in principle a free parameter. Without a loss of generality, we estimate the variance by $\langle \sigma^2\rangle_0 = 3H^4/(8\pi^2 m_{\sigma 0}^2)$,
assuming that $\sigma$ is initially in an equilibrium state with a mass $m_\sigma^2 = m_{\sigma 0}^2$ \cite{Starobinsky:1994bd}.

To see the evolution of $\rho_1$ in the standard hybrid inflation scenario \cite{Linde:1993cn}, let us firstly turn off the centrifugal force by setting $\xi = 0$.
With an initial value $\sigma_0 \neq 0$, the numerical result in Fig. \ref{fig_standard_PDF} shows that the PDF can reach back to a normal distribution with a peak at the origin by the time of phase transition.
The PDF spreads out soon during the phase transition as $m_\sigma^2$ becomes very small.
After the phase transition the PDF equally distributes to the positive and the negative minima of the potential, which indicates the well-known domain wall problem \cite{GarciaBellido:1996qt,Linde:1993cn}.

\begin{figure}
	\begin{center}
		\includegraphics[width=140mm]{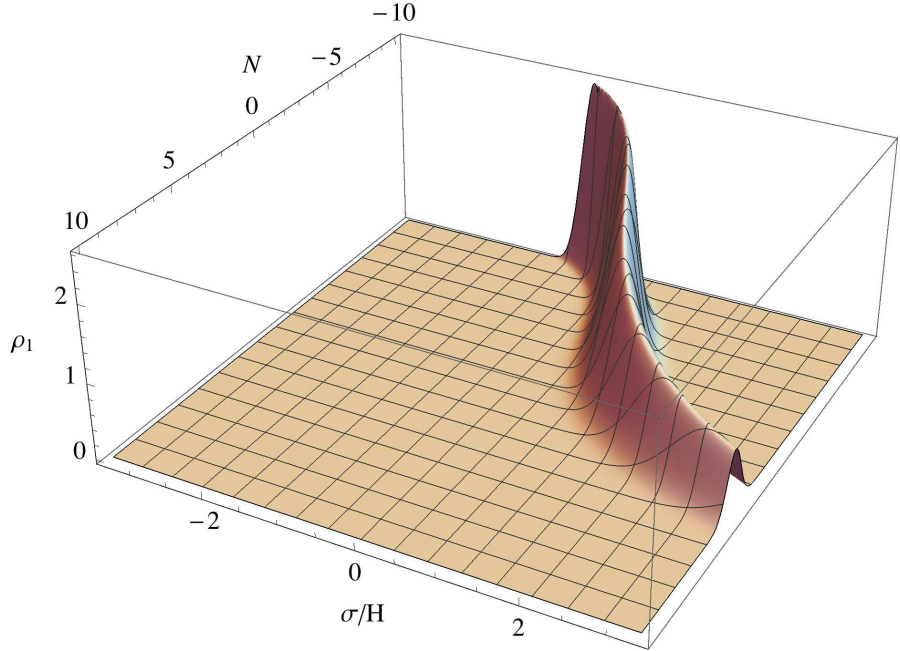}
	\end{center}
	\caption{\label{fig_turning_PDF}The evolution of the one-point probability distribution function $\rho_1$ of the waterfall field with a centrifugal force ($\xi = 0.01$). The waterfall phase transition starts at $N =0$.
		The initial value $\sigma_0 = -0.5 H$ is used.
		The potential parameters are fixed by $g/H = 4$, $\lambda = 10^{-6}$, $\theta_c = 1$, $p =1$ , $R/M_p = 10$ and $H/ M_p = 10^{-5}$.
	}
\end{figure} 

On the other hand, if we turn on the centrifugal force by taking $\xi > 0 $ the potential minimum is shifted from the origin to $\sigma > 0$ by the time of phase transition.
The numerical result in Fig. \ref{fig_turning_PDF} shows that the peak of the PDF runs into the positive region, even with an initial value $\sigma_0 < 0$, and the possibility to find $\sigma$ in the negative minimum after phase transition
can be significantly suppressed. We can estimate the probability of $\sigma$ to end up in the negative minimum by
\begin{align}
P_c = \int_{-\infty}^{0} \rho_{1 c}[\sigma]\; \frac{d\sigma}{H},
\end{align}
where $\rho_{1\, c}$ is the one-point PDF at the critical time. To evade the domain wall problem, we require $P_c < 1/ e^{3\Delta N}$ 
with $\Delta N$ being the number of e-folds required to solve the horizon problem substracted by that elapsed after the phase transition. 
We have checked that the condition $P_c < 1/ e^{3\Delta N}$ can be satisfied for $\sigma_0 = 0$, $\xi \in \{10^{-3},1 \}$ and $\Delta N \sim \mathcal{O}(40)$.
This implies that hybrid inflation with a turning trajectory is free from the domain wall problem for a wide range of initial conditions.

\section{Approaches for quantum fluctuations}\label{quantumfluctuation}

We resolve the quantum fluctuations of the inflaton ($\delta\theta = \delta\phi/R$) and the waterfall fields ($\delta\sigma$) by virtue of the equation of motion (EoM) approach \cite{Chen:2015dga},
which is in particular convenient for numerical computations. The reason is twofold. First of all, so far there is no exact solution found for the mode function
of a waterfall field, $\delta\sigma_{\bf k}$, in the hybrid inflation scenario with a time-varying mass. 
Although some analytical expressions for $\delta\sigma_{\bf k}$ can be obtained via the WKB method \cite{Nagasawa:1991zr,Salopek:1988qh} 
or the Hankel function solution after specifying
the inflaton dynamics \cite{Gong:2010zf}, these approximations capture only the dominant contributions on limited physical scales during inflation
and they might not be very useful to our purpose since the quantum clock signals are sometimes generated by subdominant parts of the mode functions
\cite{An:2017hlx,Chen:2015lza}.

Secondly, the system inevitably becomes non-perturbative in the asymptotic future ($N \gg N_c$) where 
the inflaton-isocurvaton coupling $g^2 \theta^2 \sigma^2$ gives the most important contribution to the field perturbations.
To be more explicitly, let us write down the quadratic Lagrangian for the field perturbations into
\begin{align}\label{def:L2}
\mathcal{L}_2 &= \frac{a^3}{2} R^2 \left[ (1+\bar{\sigma})^2 \delta\dot{\theta}^2 
-\frac{1}{a^2}(1+\bar{\sigma})^2 (\partial_i \delta\theta)^2 +\delta\dot{\bar{\sigma}}^2 -\frac{1}{a^2}(\partial_i\delta\bar{\sigma})^2
-(V_{\sigma\sigma} - \dot{\theta}_0^2)\delta\bar{\sigma}^2\right], \\ \label{def:dL2}
\delta \mathcal{L}_2  &= 2 a^3 R^2 (1+\bar{\sigma})\dot{\theta}_0 \delta\bar{\sigma} \delta\dot{\theta} -a^3 RV_{\theta\sigma} \delta\bar{\sigma}\delta\theta,
\end{align}
where $\delta\theta$ and $\delta\bar{\sigma}\equiv \delta\sigma/R$ are decoupled in $\mathcal{L}_2$ but mixed in $\delta \mathcal{L}_2 $.

In the interaction picture, $\mathcal{L}_2$ is used to define the EoM of ``free fields'' and $\delta \mathcal{L}_2 $ is generally cast into
 interactions, if it can be expanded perturbatively. This is one of the underlying principles for the conventional in-in formalism \cite{Weinberg:2005vy,Chen:2010xka,Wang:2013eqj}.
In the parameter regime we are interested, however, $\delta \mathcal{L}_2 $ also plays a fundamental role in the EoM of field perturbations,
and thus the definition of free fields based on $\mathcal{L}_2$ becomes insufficient. 

\begin{figure}
	\begin{center}
		\includegraphics[width=68mm]{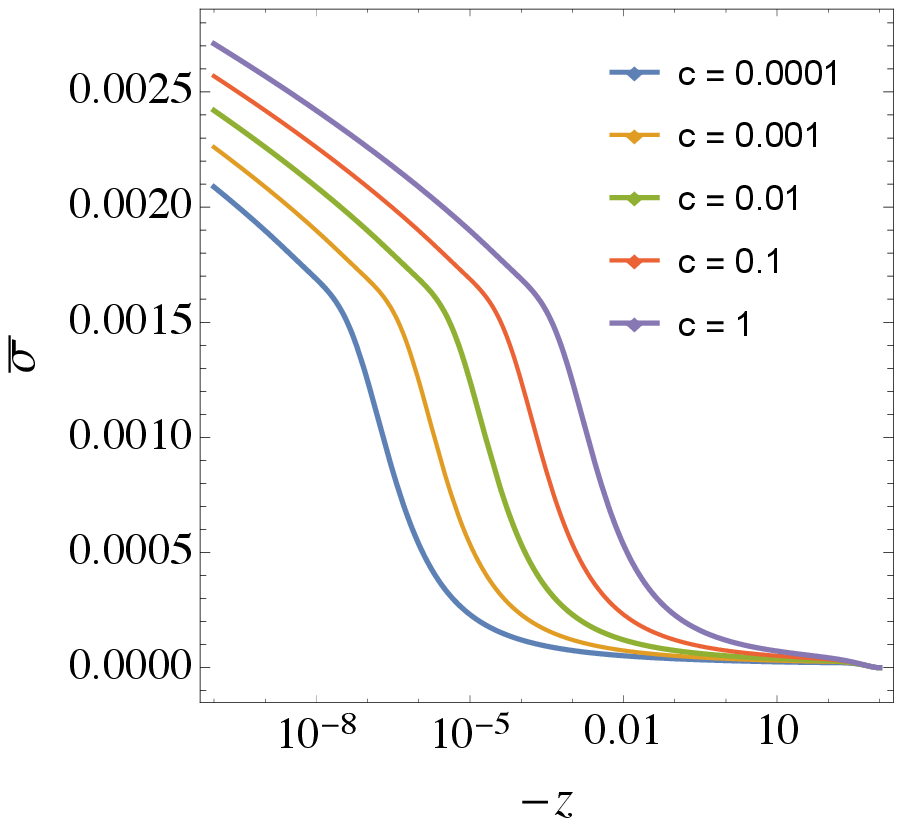}
		\hfill
		\includegraphics[width=64mm]{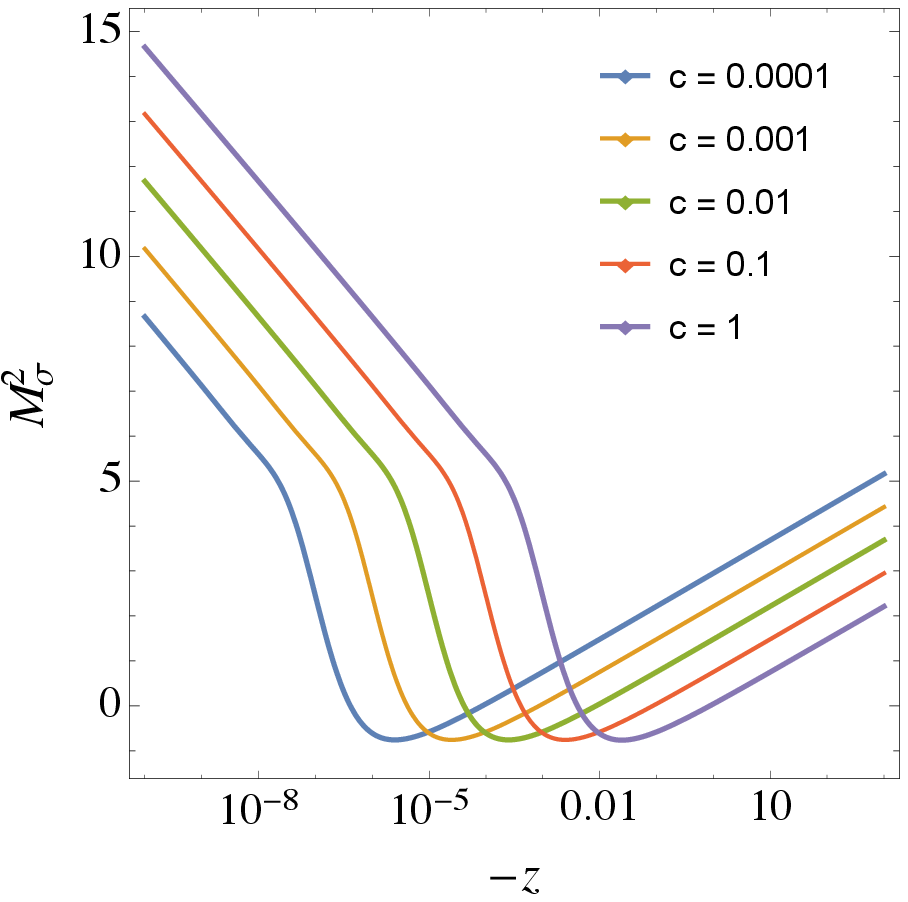}
	\end{center}
	\caption{\label{fig 3}The evolution of $\bar{\sigma}$ (left panel) and $M_\sigma^2$ (right panel) with respect to the co-moving conformal time $z$
		of various $c = k/k_c$. The parameters are fixed by $\tilde{g} \equiv g/H = 4$, $\lambda = 10^{-6}$, $\theta_c = 1$, $p =1$ , $R/M_p = 10$ and $H/ M_p = 10^{-5}$.
	}
\end{figure}

In the EoM approach \cite{Chen:2015dga}, one can define a mixed vacuum state shared by many quantum fields to deal with a strongly coupled system.
We therefore define the mode functions with respect to two sets of vacuum states as
\begin{align} 
\label{mode function: dtheta}
\delta\theta_{\bf k} &= 
u^{+}_{ k} \hat{a}_{\bf k} + u^{+\ast}_{ k} \hat{a}^\dagger_{\bf -k} + u^{-}_{ k}\hat{b}_{\bf k} +u^{-\ast}_{ k} \hat{b}^\dagger_{\bf -k}, \\
\label{mode function: dsigma}
\delta\bar{\sigma}_{\bf k} & =
v^{+}_{ k} \hat{a}_{\bf k} + v^{+\ast}_{ k} \hat{a}^\dagger_{\bf -k} + v^{-}_{ k}\hat{b}_{\bf k} +v^{-\ast}_{ k} \hat{b}^\dagger_{\bf -k}, 
\end{align}
where $\hat{a}$ and $\hat{b}$ are shared by $\delta\theta$ and $\delta\bar{\sigma}$, and they shall satisfy the commutation relations 
given by \cite{Chen:2015dga} as
\begin{align}
[\hat{a}_{\bf k}, \hat{a}^\dagger_{\bf -p}] &= [\hat{b}_{\bf k}, \hat{b}^\dagger_{\bf -p}] 
= (2\pi)^3 \delta^3( {\bf k + p}), \\
[\hat{a}_{\bf k}, \hat{a}_{\bf -p}] &= [\hat{b}_{\bf k}, \hat{b}_{\bf -p}] 
= [\hat{a}_{\bf k}, \hat{b}_{\bf -p}] = [\hat{a}^\dagger_{\bf k}, \hat{b}^\dagger_{\bf -p}]  
= [\hat{a}_{\bf k}, \hat{b}^\dagger_{\bf -p}] = [\hat{b}_{\bf k}, \hat{a}^\dagger_{\bf -p}] = 0.
\end{align}
Here $\delta\theta$ and $\delta\bar{\sigma}$ are governed by the EoMs derived from $\mathcal{L}_2$ together with $\delta \mathcal{L}_2$, which read
\begin{align}\label{eom:dtheta}
\delta\theta_k^{\prime\prime} - & \left(\frac{2}{z} - \frac{2\bar{\sigma}^\prime}{1 + \bar{\sigma}}\right) \delta\theta_k^{\prime} + \delta \theta_k 
\\\nonumber
 &\qquad = \left[\frac{2V_\theta}{R^2 H^2 (1+\bar{\sigma})} - 2\tilde{g}^2\frac{\theta\bar{\sigma}}{(1+\bar{\sigma})^2}\right] \frac{\delta\bar{\sigma}_k}{z}
 + \frac{2\sqrt{3B}}{1+ \bar{\sigma}} \left[\frac{\bar{\sigma}^\prime}{1+\bar{\sigma}}\frac{\delta\bar{\sigma}_k}{z} + \frac{\delta\bar{\sigma}^\prime_k}{z}\right], 
 \\\nonumber \label{eom:dsigma}
 \delta\bar{\sigma}_k^{\prime\prime} - &\frac{2}{z} \delta\bar{\sigma}_k^\prime  + \delta\bar{\sigma}_k + 
 \left[3\lambda \frac{R^2}{H^2}\bar{\sigma}^2 + 2 p \tilde{g}^2 \frac{M_p^2}{R^2}\ln\left(-\frac{k_c}{k}z\right) -3B \right] \frac{\delta\bar{\sigma}_k}{z^2} \\
 &\qquad = -2(1+\bar{\sigma})\sqrt{3B} \, \frac{\delta\theta^\prime_k}{z} -2 \tilde{g}^2 \theta\,\bar{\sigma}\, \frac{\delta\theta_k}{z^2},
\end{align}
where $z = k \eta$ is the conformal time co-moving with each mode $k$, and $k_c$ is the horizon scale at the time of the phase transition.
We define a dimensionless parameter $\tilde{g}\equiv g/H$.
The waterfall phase transition at $N = - \ln(-k_c z/k) = -\ln (-k_c \eta)= 0$ breaks the time-translation symmetry 
$z \rightarrow c z$ of the mode functions, where $c = k/k_c$.
Therefore $k$-dependence appears in the effective mass
\begin{align}\label{eq:mass_dsigma}
M_\sigma^2 (z) = 3\lambda \frac{R^2}{H^2}\bar{\sigma}^2 + 3A \ln\left(-\frac{z}{c}\right) -3B, 
\end{align}
when using $z$ as the time parameter, but the scale-dependence in $M_\sigma^2$ disappears if we use the usual conformal time $\eta = z/k$.
Similarly, the EoM of the background field given by 
\begin{align}\label{eom:sigma_bar in z}
\bar{\sigma}^{\prime\prime} -\frac{2}{z} \bar{\sigma}^\prime + 
\left[\lambda\frac{R^2}{H^2}\bar{\sigma}^2 + 3A \ln\left(-\frac{z}{c}\right)\right] \frac{\bar{\sigma}}{z^2} = \frac{3B}{z^2},
\end{align}
looks different for each $k$ mode in its co-moving time $z$.
As shown in Fig. \ref{fig 3}, modes with $k < k_c$ (or $c < 1$) experience the waterfall transition at $-z = c$ after they have exited the horizon at $-z = 1$.
The $k$-dependence in $\bar{\sigma}$ and $M_\sigma^2$ simply reflects the time dependence of the background solution in the evolution of perturbations.

The only requirement for solving the field perturbations by 
the EoM approach is that $\delta\theta$ and $\delta\sigma$ shall satisfy a well-posed initial condition \cite{Chen:2015dga}.
Fortunately in the limit well inside the horizon ($-z \rightarrow \infty$) Eq. \eqref{eom:dtheta} and Eq. \eqref{eom:dsigma} simply reduce to
\begin{align}
\delta\theta_k^{\prime\prime} + \delta \theta_k =0, \qquad \mathrm{and}\qquad  \delta\bar{\sigma}_k^{\prime\prime} + \delta\bar{\sigma}_k  =0,
\end{align}
which implies the solution of $\delta\theta \propto e^{-i z}$ and $\delta\sigma \propto e^{-i z}$.
After imposing the initial condition $\bar{\sigma} = 0$, the EoM \eqref{eom:dtheta} and \eqref{eom:dsigma} coincide with the equations of the
non-perturbative quasi-single field inflation \cite{An:2017hlx,An:2017rwo} up to the order of $1/z$.
Thus the initial mode functions are found as
\begin{align}\label{eq:initial mode function}
u_k^\pm = \frac{H}{R\sqrt{4k^3}}e^{-iz} (-z)^{1\pm i p M_p^2/(\theta_c R^2) }, \qquad  v_k^\pm = \pm i u_k^\pm. 
\end{align}
The slow-roll condition $M_p^2 \ll R^2$ for the potential \eqref{V_choatic} implies that the two sets of mode function 
$u^\pm_k$ (and thus $v_k^\pm$) are almost degenerate up to a phase factor.

\section{Imprints of waterfall phase transition}\label{Sec:Imprints}

\subsection{Power spectrum}
We now solve the correlation functions of $\delta\theta$ and $\delta\sigma$ through the EoMs \eqref{eom:dtheta}, \eqref{eom:dsigma} 
and \eqref{eom:sigma_bar in z} from the initial states \eqref{eq:initial mode function}.
The two-point correlation function for field perturbations based on the mixed vacuum states \eqref{mode function: dtheta} and \eqref{mode function: dsigma}
are defined as \cite{Chen:2015dga}:
\begin{align}\label{eq:2-point theta}
\left\langle \delta\theta_{\bf k} (z) \delta\theta_{\bf p} (z) \right\rangle =
(2\pi)^3 \delta^3({\bf k + p}) \left[u^+_{  k}(z) u^{+ \ast}_{  p} (z) + u^-_{  k}(z) u^{-\ast}_{  p}(z) \right],
\end{align}
and for $\langle \delta\sigma^2\rangle$ one simply replaces $u_k^\pm$ by $v_k^\pm$, respectively. 
In the interaction picture, the definition \eqref{eq:2-point theta} includes 
the resummed tree-level corrections from $\delta \mathcal{L}_2 $ to the free $\delta\theta$ field defined by $\mathcal{L}_2 $.
The gauge invariant curvature perturbation is related to \eqref{eq:2-point theta} by \cite{Chen:2009zp}
\begin{align}
\left\langle \zeta_{\bf k} \zeta_{\bf p} \right\rangle = \frac{H^2}{\dot{\theta}^2} \left\langle \delta\theta_{\bf k} \delta\theta_{\bf p}  \right\rangle
= (2\pi)^3 \delta^3({\bf k + p}) \frac{H^2}{\dot{\theta}^2} \frac{2\pi^2}{k^3} P_\theta (k).
\end{align} 

\begin{figure}
	\begin{center}
		\includegraphics[width=68mm]{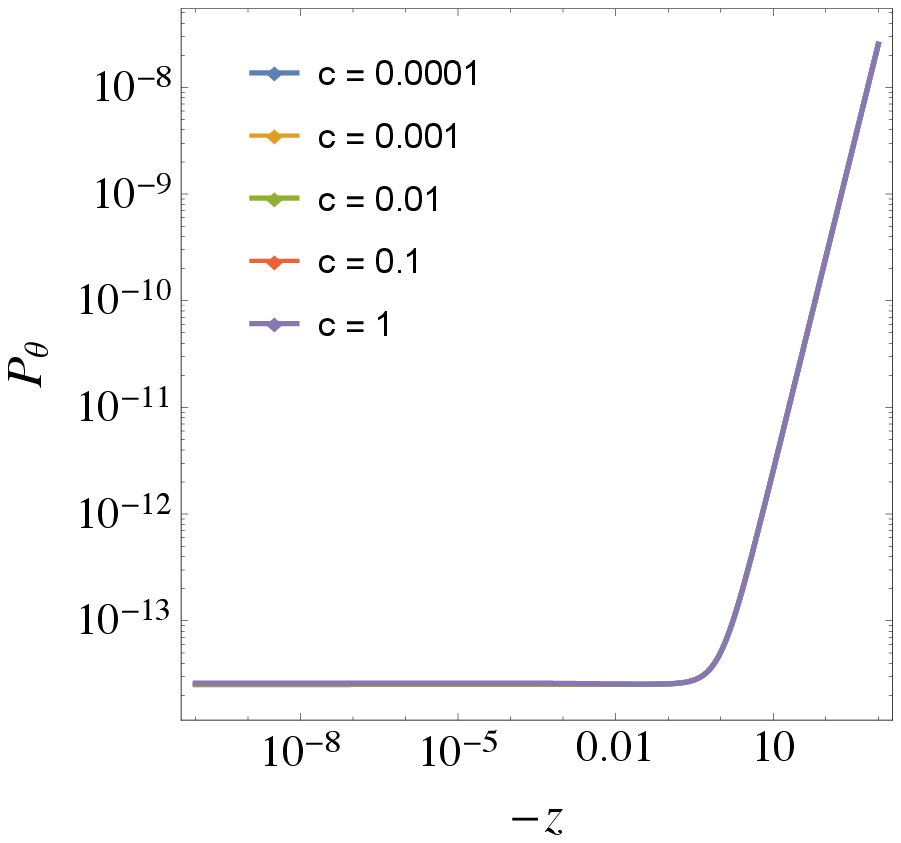}
		\hfill
		\includegraphics[width=68mm]{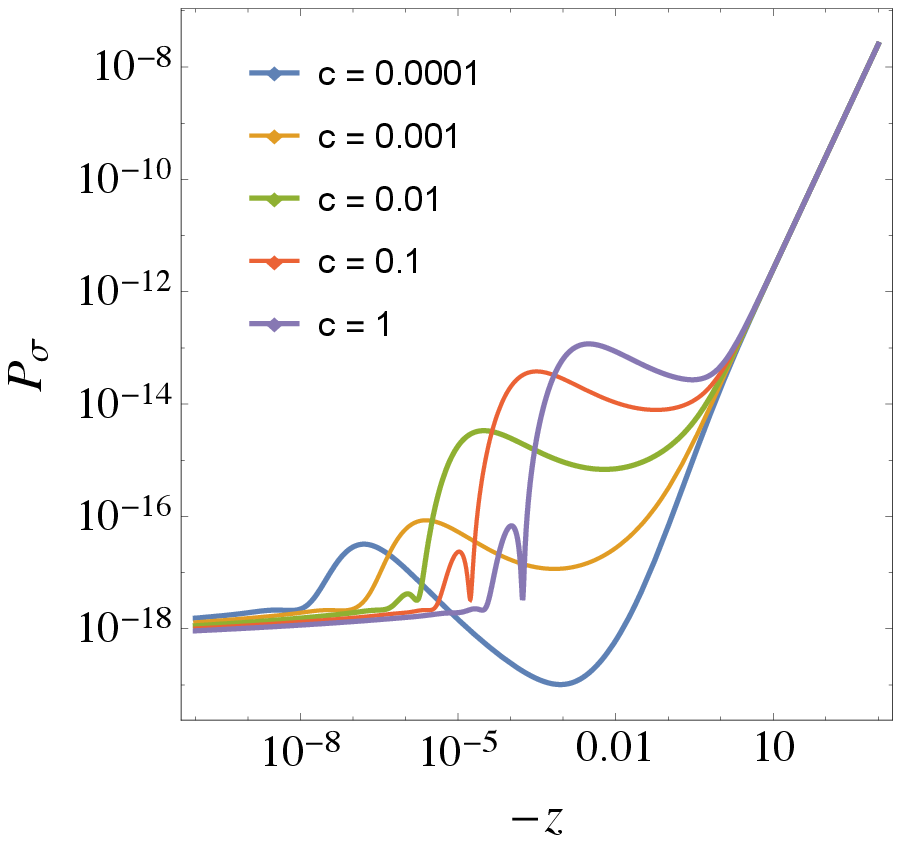}
	\end{center}
	\caption{\label{fig 4}The evolution of $P_\theta(c,z)$ (left panel) and $P_\sigma(c, z)$ (right panel) with respect to the co-moving conformal time $z$
		of various $c = k/k_c$. The parameters are fixed by $\tilde{g} \equiv g/H = 4$, $\lambda = 10^{-6}$, $\theta_c = 1$, $p =1$ , $R/M_p = 10$ and $H/ M_p = 10^{-5}$.
	}
\end{figure}

The results of Fig. \ref{fig 4} indicate that $\delta\theta$ is governed by two kinds of solutions.
One can check that $\delta\theta$ decays with a power of $-z$ inside the horizon ($-z > 1$),
since we have chosen a small kinetic coupling between $\theta$ and $\sigma$, 
and it goes to a constant on superhorizon scales ($-z < 1$).
On the other hand, the evolution of $\delta\sigma$ essentially has three phases.
While $\delta\sigma$ remains well inside the horizon ($-z \gg 1$), it is governed by the initial state solution \eqref{eq:initial mode function}
scaling as $-z$. Once the mass $M_\sigma^2$ becomes important as $M_\sigma^2 \simeq -z$, $\delta\sigma$ starts to oscillate.
Note that $\delta\sigma$ grows with time at the beginning of the oscillation phase since $M_\sigma$ becomes temporarily negative.
As $-z \rightarrow 0$, $\sigma$ has settled down in the new minimum (the true vacuum) given by \eqref{sigma_minima}, and 
$\delta\bar{\sigma}$ obtain a very large mass from the evolution of $\sigma_0$. In this case one may omit all derivatives of $\delta\sigma$
in \eqref{eom:dsigma} to find the mildly decaying long wavelength solution as
\begin{align}\label{eq:dsigma_IR}
\delta\bar{\sigma}^{\rm IR} \approx -2\frac{g^2}{H^2}\theta_c \frac{\bar{\sigma}_{\rm min}}{M_\sigma^2} \delta\theta^{\rm IR}. 
\end{align} 
Here $\delta\theta^{\rm IR} \approx H/(\sqrt{2}R)$ stands for the asymptotic value of $\delta\theta_k$ in the limit $-z \rightarrow 0$, which is nearly scale-independent. This kind of effective field theory approach can be applied to both background level and perturbation level as is widely used in the previous literatures \cite{Tolley:2009fg,Achucarro:2010jv,Baumann:2011su,Achucarro:2012sm,Chen:2012ge,Pi:2012gf,Achucarro:2012yr,Gwyn:2012mw,Gong:2013sma,Tong:2017iat,Iyer:2017qzw}.
By using \eqref{sigma_minima}, we find that $M_\sigma^2 \rightarrow 6 A N$ as $N \gg 1$ and therefore the asymptotic value 
behaves as $\delta\sigma_k \sim 1/\sqrt{-\ln(-z/c)}$. 

\begin{figure}
	\begin{center}
		\includegraphics[width=64mm]{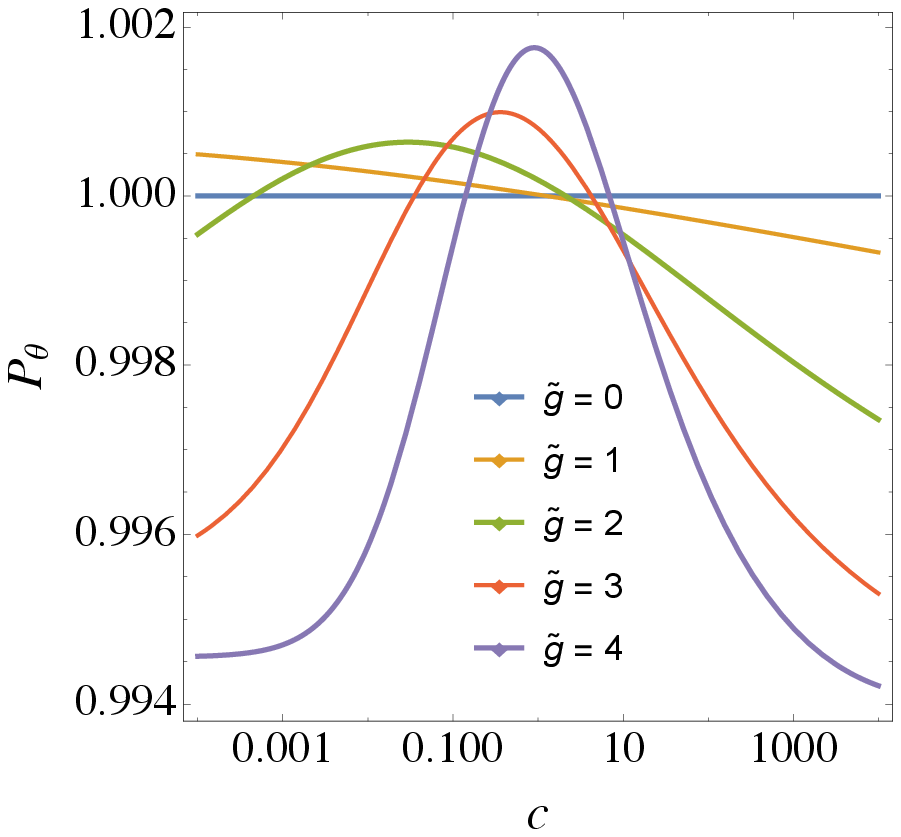}
		\hfill
		\includegraphics[width=71mm]{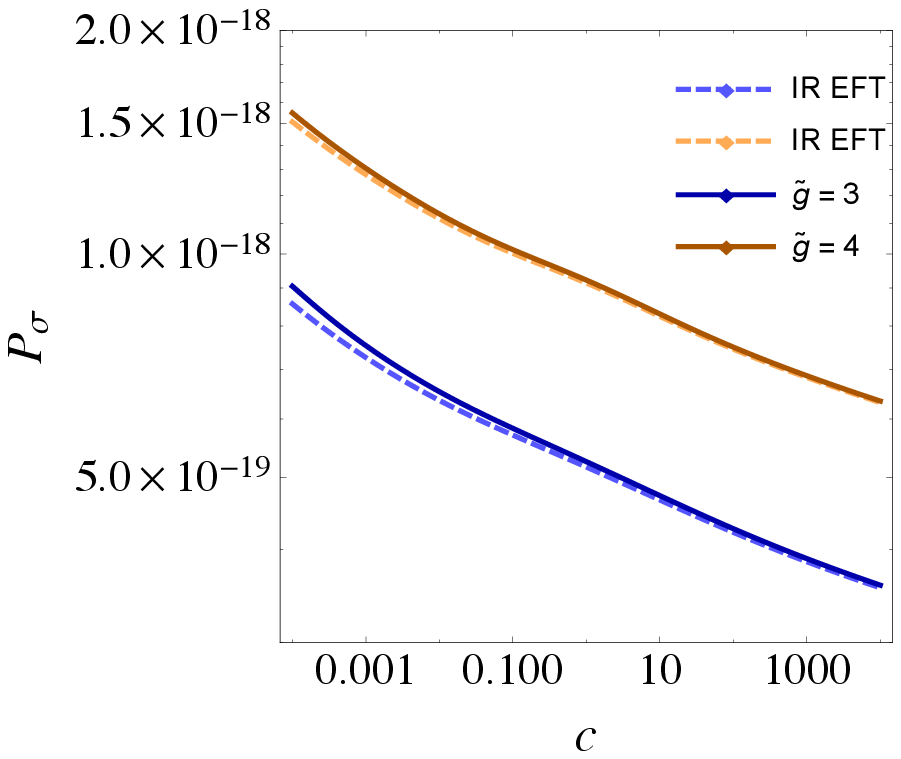}
	\end{center}
	\caption{\label{fig 5}The value of $P_\theta$ (left panel) and $P_\sigma$ (right panel) in terms of $c = k/k_c$. 
		The value of $P_\theta$ is normalized with respect to the constant turn scenario without the waterfall transition ($\tilde{g}=0$).
		The dashed lines in the right panel are given by \eqref{eq:dsigma_IR}.
		The parameters are fixed by $\lambda = 10^{-6}$, $\theta_c = 1$, $p =1$ , $R/M_p = 10$ and $H/ M_p = 10^{-5}$.
	}
\end{figure}

We examine the scale-dependence of $P_\theta$ and $P_\sigma$ in Fig. \ref{fig 5}.
The scale-invariance of $P_\theta$ is preserved in the constant turn limit $\tilde{g} = 0$ where there is no phase transition and is mildly broken with 
the increase of $\tilde{g}$. In the large $\tilde{g}$ limit $P_\theta$ peaks around $k_c$, which recovers the finding in the hybrid inflation
without a turning trajectory \cite{Abolhasani:2012px}. The scale-dependence of $P_\sigma$ is led by the heavy field approximation
\eqref{eq:dsigma_IR} in the IR regime.

\subsection{Bispectrum}

We want to search for the imprint of the waterfall field in the three-point function of the curvature perturbation $\zeta$. 
For this purpose, we pick up interactions from field perturbations at cubic order and compute their contribution to $\langle\zeta^3\rangle$
by using the in-in formalism \cite{Weinberg:2005vy,Chen:2010xka,Wang:2013eqj}. 
The underlying assumption is that these higher-order corrections are perturbatively small compared with the
strongly coupled solutions obtained in the previous section.
The most important decay channel at tree-level shall come from the $\sigma$ self-interaction \cite{Chen:2009zp,An:2017hlx,An:2017rwo}
 \begin{align}\label{def:cubic_interaction}
 \mathcal{L}_3 &= -\frac{1}{6} a^3 R^3V_{\sigma\sigma\sigma} \delta\bar{\sigma}^3 = - a^3 \lambda R^4 \bar{\sigma} \delta\bar{\sigma}^3.
 \end{align}
This self-interaction is treated perturbatively provided that $\lambda R \bar{\sigma} < H$. 
There are some other cubic interactions 
$ g^2 R^2\bar\sigma \delta \bar{\sigma }\delta \theta\delta \theta$ and $g^2 R^2\theta_c \delta\bar{\sigma} \delta\bar{\sigma}\delta\theta$
induced by the inflaton-isocurvaton coupling. Having in mind that $R \gg M_p$ and $g \sim H$, these interactions 
are less important than \eqref{def:cubic_interaction} provided that $\lambda > H^2/R^2$.  

Part of the conventional in-in formalism for the perturbative quasi-single field inflation \cite{Chen:2009zp} 
is now adjusted for a strongly coupled system \cite{An:2017hlx}.
First of all, the definition \eqref{mode function: dtheta} and \eqref{mode function: dsigma} in fact allows a direct decay of $\sigma$ into $\theta$
(and vice versa), as can be seen by the non-vanishing correlation function
\begin{align}\label{def:2pt_theta_sigma}
\left\langle \delta\theta_{\bf k}(z) \delta\sigma_{\bf p}(z_1) \right\rangle 
= (2\pi)^3 \delta^3({\bf k + p}) \left[u^+_{ k}(z) v^{+ \ast}_{ p} (z_1) + u^-_{ k}(z) v^{-\ast}_{ p}(z_1) \right],
\end{align} 
where $z_1 \equiv k_1 \eta$.
It is also straightforward to define a mixed propagator 
\begin{align}
G_R(k; z, z_1) &\equiv -i \Theta(z - z_1) \left[\delta\theta_{ k}(z), \delta\sigma_{k}(z_1)\right], \\ 
&= -\frac{i}{(2\pi)^3} \Theta(z - z_1) \left\langle \delta\theta_{\bf k}(z) \delta\sigma_{\bf p}(z_1) \right\rangle, 
\end{align}
where $G_R$ acts as the retarded Green's function for both \eqref{eom:dtheta} and \eqref{eom:dsigma}.

We then define the interaction picture field $\delta\theta_I$, $\delta\sigma_I$ 
as the solutions of the coupled equations \eqref{eom:dtheta} and \eqref{eom:dsigma}, which are evaluated by the EoM approach. 
To obtain the value of correlation functions in the asymptotic limit $-z \rightarrow 0$, it is more convenient to apply the
commutator representation of the in-in formalism \cite{Weinberg:2005vy,Chen:2010xka,Wang:2013eqj}.
At tree-level, the higher-order corrections $H_I$ to the three-point function of $\delta\theta$ is then calculated by
\begin{align}\label{eq:3point-commutator}\nonumber
\langle \delta\theta^3(z)\rangle &= \left\langle
 \left( \sum_{N=0}^{\infty}i^N \int^z dz_N \cdots \int^{z_2}dz_1 
[H_I(z_1), \cdots [H_I(z_{N}),\delta\theta^3_I(z)]] \right) \right\rangle \\
&= i \int^{z}dz_1 \left\langle 0\vert  \left[H_I(z_1), \delta\theta^3_I(z)\right] \vert 0\right\rangle .
\end{align}
By choosing $H_I(z) = \lambda R^4 a^4(z) \bar{\sigma}(z) \delta\sigma^3(z)$ from \eqref{def:cubic_interaction}, where $a(z) = -k/(Hz)$, 
the three-point function \eqref{eq:3point-commutator} reads
\begin{align}\label{eq:3point_general}
 \langle \delta\theta^3(z)\rangle
&= -2 \lambda R^4 \frac{k_1^3}{H^4}\, \mathrm{Im} 
\left[  \int^{z}\frac{dz_1}{z_1^4}\, \bar{\sigma}(\frac{z_1}{k_1}) G_R(k_1; z, z_1)G_R(k_2; z, \frac{k_2}{k_1}z_1)G_R(k_3; z, \frac{k_3}{k_1}z_1)\right].
\end{align} 
This expression can be identified as the mixed propagator approach of the Schwinger-Keldysh diagrammatics interpretation
from a set of decoupled initial states \cite{Chen:2017ryl}.

\begin{figure}
	\begin{center}
		\includegraphics[width=100mm]{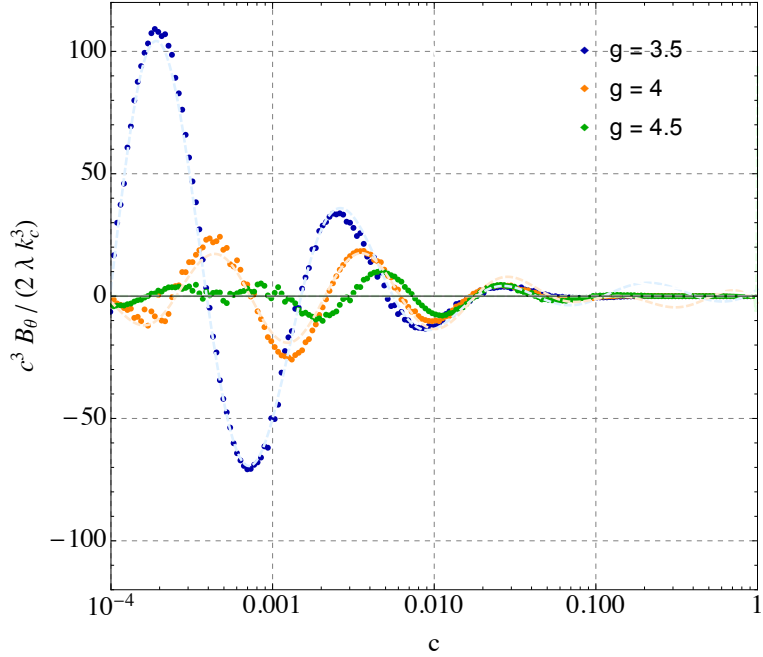}
	\end{center}
	\caption{\label{fig 6}The squeezed limit of the bispectrum $B_\theta$ as a function of $c$ with $g = \{3.5,4, 4.5\}$ in the unit of $H$.
		The dashed lines are the bestfit functions given by Eq. \eqref{eq:model fitting}.
	The parameters are fixed by $\lambda = 10^{-6}$, $\theta_c = 1$, $p =1$ , $R/M_p = 10$ and $H/ M_p = 10^{-5}$.
	}
\end{figure}

For convenience we shall fix $k_1 = k_2 = k_c$ and reparametrize $k_3 = c k_c$.
The integration \eqref{eq:3point_general} is now reduced as
\begin{align}\label{eq:3point_squeezed}
 \langle \delta\theta^3(z)\rangle = -2 \lambda R^4 \frac{k_c^3}{H^4}\, \mathrm{Im}
 \left[ \int^{z}\frac{dz_1}{z_1^4}\, \bar{\sigma}(\frac{z_1}{k_c}) G_R^2(k_c; z, z_1) G_R(c k_c; z, c z_1)\right],
\end{align}
where some of the numerical results are given in Fig. \ref{fig 6},
and we have used the definition for the bispectrum $B_\theta$ as
\begin{align}
\langle \delta\theta_{\bf k_1}(z)\delta\theta_{\bf k_2}(z)&\delta\theta_{\bf k_3}(z) \rangle 
\equiv (2\pi )^4 \delta^3({\bf k_1+ k_2 + k_3}) B_\theta({\bf k_1, k_2 ,k_3}).
\end{align}

Some of the numerical results of $B_\theta$ are given in Fig. \ref{fig 6}. We have chosen the coupling constant $g > 3 H$
so that the $\sigma$ field can reach the new minimum in time for a given IR cutoff $z= z_f$. 
The numerical results with $g = 3.5 H$ in Fig. \ref{fig 6} exhibit a similar shape as the oscillatory component of the bispectrum 
from the constant turn inflation \cite{Chen:2015lza,Chen:2017ryl}. It is amplified in the direction towards the squeezed limit ($c \rightarrow 0$).
With the existence of a waterfall phase transition, however, the oscillation component of $B_\theta$ 
 becomes dominantly large relative to the non-oscillatory contribution.

One can compare the results of $B_\theta$ with the bispectrum shape function $S_\theta$ used in \cite{Chen:2015lza,Chen:2017ryl} through the relation $B_\theta = (2\pi)^3 P_\theta^2 S_\theta/ ({ k_1 k_2 k_3})^2$.
Then the result in Fig. \ref{fig 6} is proportional to $c^3 B_\theta \propto c\, S_\theta$.
In the equilateral limit where $c \rightarrow 1$, the results can be nicely fitted by the analytical form
\begin{align} \label{eq:model fitting}
\frac{c^3}{2\lambda k_c^3}B_\theta (c) = \alpha_1 c^{\beta_1}\cos\left(\omega_1 \ln c - \varphi_1\right) +
\alpha_2 c^{\beta_2} \sin\left(\omega_2 \ln c - \varphi_2\right) + \alpha_3 c^{2},
\end{align}
based on the qualitative analysis constructed in \cite{An:2017hlx,An:2017rwo}. 
The term $\alpha_3 c^{2}$ comes from the decaying solution in $\delta\sigma$, which is negligible in the squeezed limit.
The parameters of the oscillatory terms depend on the mass $m_\sigma^2$.
For a constant mass $m_\sigma^2 >0$ one shall find $\beta_1 = \beta_2 = 3/2$ and $\omega_1 = \omega_2$.
Here we have the fitting results $\{\beta_1, \beta_2, \omega_1, \omega_2 \} = \{ -0.88, -0.88 , 2.42, 2.49\}$ for $g = 3.5$ and
$\{\beta_1, \beta_2, \omega_1, \omega_2\} = \{-0.39, -0.44, 3.39, -2.49\}$ for $g =4$.

The results with $g \geq 4 H$ are neither amplified in the squeezed limit nor in the equilateral limit,
and therefore they cannot be fitted by the analytical function \eqref{eq:model fitting}. 
In this case a novel shape of the oscillatory bispectrum is generated, with a maximal amplitude in between the squeezed and the equilateral limits.
This result is easy to be understood since the effective mass of the isocurvaton perturbation \eqref{eq:mass_dsigma} is changing from a large value
at $-z \gg c$ to be very small around $-z = c$ and then becomes large again as $z \rightarrow z_f$.
In the equilateral limit where $c < 0.01$, the shape of the bispectrum with $g \geq 4 H$ is the same as the cases with a small coupling constant $g$.
For a large coupling $g$, the mode function $\delta\sigma_k$ with $k \ll k_c$ can decay to be much smaller than the asymptotic value
$ \delta\sigma^{\rm IR}$ given by \eqref{eq:dsigma_IR}, and it may not reach back to $ \delta\sigma^{\rm IR}$ in time.
As a result, the contribution from modes $\delta\sigma_k$ with $k \rightarrow 0$ are suppressed in the integration \eqref{eq:3point_squeezed},
consistent with the finding from the constant turn scenario with a large kinetic coupling case in \cite{An:2017hlx,An:2017rwo}. 

We have checked that the oscillation amplitude with $g < 4H$ also starts to decay as we extend the IR cutoff to $c<10^{-4}$.
This means that the phase transition occurs early enough such that the isocurvaton can find its new minimum in time even with a small $g$
(see Fig. \ref{Fig2}). We thus conclude that, as long as $-z/c$ is small enough, 
the result of $c^3 B_\theta$ always converges in the squeezed limit for arbitrary values of $g/H > 0$. 
Nevertheless for $g^2/H^2 < 1$ one can treat $\delta\mathcal{L}_2$ given by \eqref{def:dL2} as perturbative interactions and proceed all the computation
by the canonical in-in formalism, as shown in \cite{Chen:2009zp}.

We note that the shape of $B_\theta$ given by $g = 3.5 H$ could be what is really observed in this scenario,
provided that there is enough time for all the superhorizon modes to reach the asymptotic value at the IR cutoff for observations.

\section{Discussions}\label{Sec:discussion}
Particles with masses around the size of the Hubble scale during inflation are important targets for the cosmological collider research.
In this work we investigate the characteristic signals from a class of scalar particles that experience of a waterfall phase transition
according to the mechanism introduced by the hybrid inflation \cite{Linde:1993cn}. As an explicit study case,
we incorporate the waterfall potential with the constant turn inflation \cite{Chen:2009zp}, and we restrict the parameter space to that of the
quasi-single field regime where the energy density of the isocurvaton is always subdominant.

With the help of the centrifugal force produced by the turning potential valley, the isocurvaton can obtain non-trivial initial conditions at the critical time for
the phase transition, depending on the model parameters. We numerically solved the background equation of motion and found that the 
waterfall phase transition can last for more than 10 e-folds with a sufficiently small coupling constant $\tilde{g} =g/H$. In the limit of $g \rightarrow 0$, 
$\theta_c \rightarrow \infty$ and there is no phase transition occur. Thus we are interested in the case with $g \geq H$.

We have firstly checked the behavior of the long wavelength modes of the isocurvature field under the waterfall phase transition. The long wavelength part of the waterfall field is treated as an auxiliary classical field governed by the stochastic process \cite{Starobinsky:1986,Starobinsky:1994bd}.
We found that hybrid inflation in a turning potential valley can avoid the domain-wall formation problem from a wide range of initial conditions.
We also confirmed that the decomposition of the waterfall field $\sigma$ into a classical part $\sigma_0$ and a quantum-fluctuation part $\delta\sigma$ is generically well-posed in this scenario.

We have resolved the quantum fluctuations $\delta\sigma$ by virtue of equation of motion approach \cite{Chen:2015dga} base on the modified in-in formalism
for a strongly coupled system. We applied the EoM method with mixed initial states to compute the two-point and three-point correlation functions.
The appearance of a spiky feature in the power spectrum with a large $g$ shows that the time-translation invariance
of the field perturbations in the IR limit is explicitly broken by the waterfall phase transition.


Thanks to the waterfall phase transition, 
our numerical results show that the oscillatory component in the bispectrum overwhelms the non-oscillatory contributions.
One may obtain two kinds of oscillating shape in this scenario. 
The first kind is a shape with an increasing oscillating amplitude in the direction of the squeezed limit, 
corresponding to the case that the IR cutoff for observations is taken at the time where all the isocurvaton field perturbations can reach
 the asymptotic value $ \delta\sigma^{\rm IR}$ given by \eqref{eq:dsigma_IR}.
This shape is also demonstrated in \cite{Chen:2015lza,Chen:2017ryl} for the constant turn inflation, but with an additional magnification factor supplied.
The second kind is a new shape of bispectrum generated where the superhorizon modes in the limit $k \rightarrow 0$ are overdamped by the time of our observation.
In this case the maximal oscillating amplitude occurs in the intermediate region between the squeezed limit and the equilateral limit.
We shall remark that the increase of the oscillating amplitude from the equilateral limit in both cases are generated even if the
effective mass $M_\sigma^2$ given by \eqref{eq:mass_dsigma} is always positive.

Finally, we want to emphasize that the purpose of the current study is to demonstrate the existence of a novel oscillatory bispectrum which could be taken as the signature of a waterfall phase transition during inflation.
One may argue from the results of Fig. \ref{fig 5} that the parameters with $g/H > 3$, used as our examples for the quantum clock signals in Fig. \ref{fig 6}, are in fact ruled out by current observations if the critical scale $k_c$ locates in the CMB region (namely $0.008 {\rm Mpc}^{-1} < k_c < 0.1 {\rm Mpc}^{-1}$).
A robust constraint of the model parameters with observational data has been postponed as a future effort.
It could be also interesting to study the loop corrections in this scenario, although the EoM approach for quantum fluctuations has effectively resummed all tree-level contributions from both perturbative and non-perturbative interactions, 
given that loop corrections in a smooth phase transition involved with tachyonic instability are found to be important if quantum fluctuations become the dominant part of the isocurvature fields~\cite{Wu:2017lnh}.

\section*{acknowledgments}
We thank Shi Pi, Teruaki Suyama and Rong-Fu Wu for several inspiring discussions, and we appreciate the technical support from Xi Tong and Louis Yang.
YW acknowledges the International Symposium on Cosmology and Particle Astrophysics (CosPA) 2017 conference held at Yukawa Institute for Theoretical Physics, Kyoto University (YITP-W-17-15), during which part of the work was accomplished.
YW is supported by ECS Grant 26300316 and GRF Grant 16301917 from the Research Grants Council of Hong Kong.
YPW is supported by JSPS International Research Fellows and JSPS KAKENHI Grant-in-Aid for Scientific Research No. 17F17322.
JY was supported by JSPS KAKENHI Grant-in-Aid for Scientific Research No. 15H02082, and Grant-in-Aid for Scientific Research
on Innovative Areas No. 15H05888.
SZ is supported by the the Hong Kong PhD Fellowship Scheme (HKPFS) issued by the Research Grants Council (RGC) of Hong Kong.

\end{document}